\documentclass[sigconf]{acmart}
\usepackage{graphicx} % Required for inserting images
\usepackage{paralist}
\usepackage{tabularray} % in the preamble
\usepackage{makecell} % in the preamble
\acmConference[ICSE-SERS '26]{2026 IEEE/ACM 48th International Conference on Software Engineering}{April 14, 2026}{Rio de Janeiro, Brazil}

\title{Enhancing Understandability and Transparency of Research Software: Tracing Research to Code}

\author{Adrian Bajraktari}
\affiliation{%
  \institution{University of Cologne}
  \city{Cologne}
  \country{Germany}}
\email{bajraktari@cs.uni-koeln.de}
\orcid{0000-0001-5997-7156}

\author{Andreas Vogelsang}
\affiliation{%
  \institution{University of Duisburg-Essen}
  \city{Essen}
  \country{Germany}}
\email{andreas.vogelsang@uni-due.de}
\orcid{0000-0003-1041-0815}

\begin{document}

\begin{abstract}
    Modern research heavily relies on software. A significant challenge researchers face is understanding the complex software used in specific research fields. We target two scenarios in this context, namely long onboarding times for newcomers and conference reviewers evaluating replication packages. We hypothesize that both scenarios can be significantly improved when there is a clear link between the paper's ideas and the code that implements them. As a time- and staff-saving approach, we propose an LLM-based automation tool that takes in a paper and the software implementing the paper, and generates a trace mapping between research ideas and their locations in code. Initial experiments have shown that the tool can generate quite useful mappings.
\end{abstract}

\maketitle

\section{Introduction}
Software has not only become omnipresent in industry, society, and everyday life, but also in research~\cite{felderer2020}. Researchers nowadays heavily rely on software for their studies, experiments, simulations, and evaluations. In a survey~\cite{hettrik14} already conducted in 2014, more than 90\% of participants reported relying on software in some way, and more than 60\% reported they could not conduct further research in their field without software support.

Software in research spans a wide range, from small one-shot scripts for a single paper to large infrastructures that last for several years and act as the backbone of the group's research activities. The latter often introduces complexity. The projects are just too big to grasp in a workday or two. In a previous study~\cite{rePaper}, researchers reported that understanding the existing codebase first before producing original research takes a long time, especially for newcomers, e.g., new PhD students or bachelor's/master's theses, resulting in months of up-front preparation. These long times stem not only from the complexity of the projects themselves, but also from the fact that experienced researchers do not have enough time to guide newcomers through the projects. Thus, new scientific staff are left alone with the code, existing papers, and, at best, somewhat decent documentation.

Another scenario in which understanding research software plays a crucial role, applying to large and small software alike, is faced by reviewers and artifact evaluators at conferences. Their role is to check whether the code provided in a replication package realizes what is stated in the paper. Besides many other obstacles~\cite{winter2022} that make the process complex and lengthy, they are entirely unfamiliar with the code and the research presented in the paper, and have no one to ask about it. For a thorough software evaluation, reviewers should ideally verify that the code implements all claims, ideas, methods, simulations, and other aspects described in the paper. This is a time-consuming and challenging task, given the problems described before.

While both scenarios affect different groups of researchers at various levels, the core problem of both is that it is hard to understand research software if you have not been heavily involved for a long time. To this end, we propose an LLM-based approach that extracts research ideas from papers and establishes trace links between these ideas and locations in associated code. With this, we aim to support understanding and transparency of research software, especially for the two scenarios mentioned above.

\section{Approach}
Our LLM-based tool takes the paper's PDF and a git repository link as inputs and executes the following process steps:
\begin{compactenum}
    \item \textbf{Preprocess}: pull the repository and transform the PDF of the paper into a string-based format\footnote{Using PyPDF2} that can be passed through the prompt to the LLM.
    \item \textbf{Create natural language representations (NLR) for each code block}. Code bases often exceed the maximum context window size for LLMs. Thus, we first let the LLM create natural language summaries, which, in comparison, use fewer tokens for each ``code block''\footnote{a file, a class, a method, a cell in a Jupyter notebook, or a logically cohesive group of code lines, e.g., in larger Python scripts with no real code structure}. 
    \item \textbf{Extract all research concepts from the paper}. A research concept is anything stated in the paper that should also appear somewhere in the code, e.g., data preprocessing, experiments, calculations, methods, results, problems, insights, claims, etc., but original to the paper or a relevant basis needed for the paper's contribution.
    \item \textbf{Create a trace link map between research concepts and code locations}. For each research concept, the LLM returns all code locations that contribute to its implementation.
\end{compactenum}
For our experiments, we used GPT-4o to handle all three major steps (2--4). Steps 2 and 4 are done in one request to the LLM's API each. Because of the size of many software repositories, transforming code into NLRs is not possible within a single context window, so the code is split into large chunks, with each chunk transformed in a separate request. 

\begin{figure}
    \centering
    \includegraphics[width=1\linewidth]{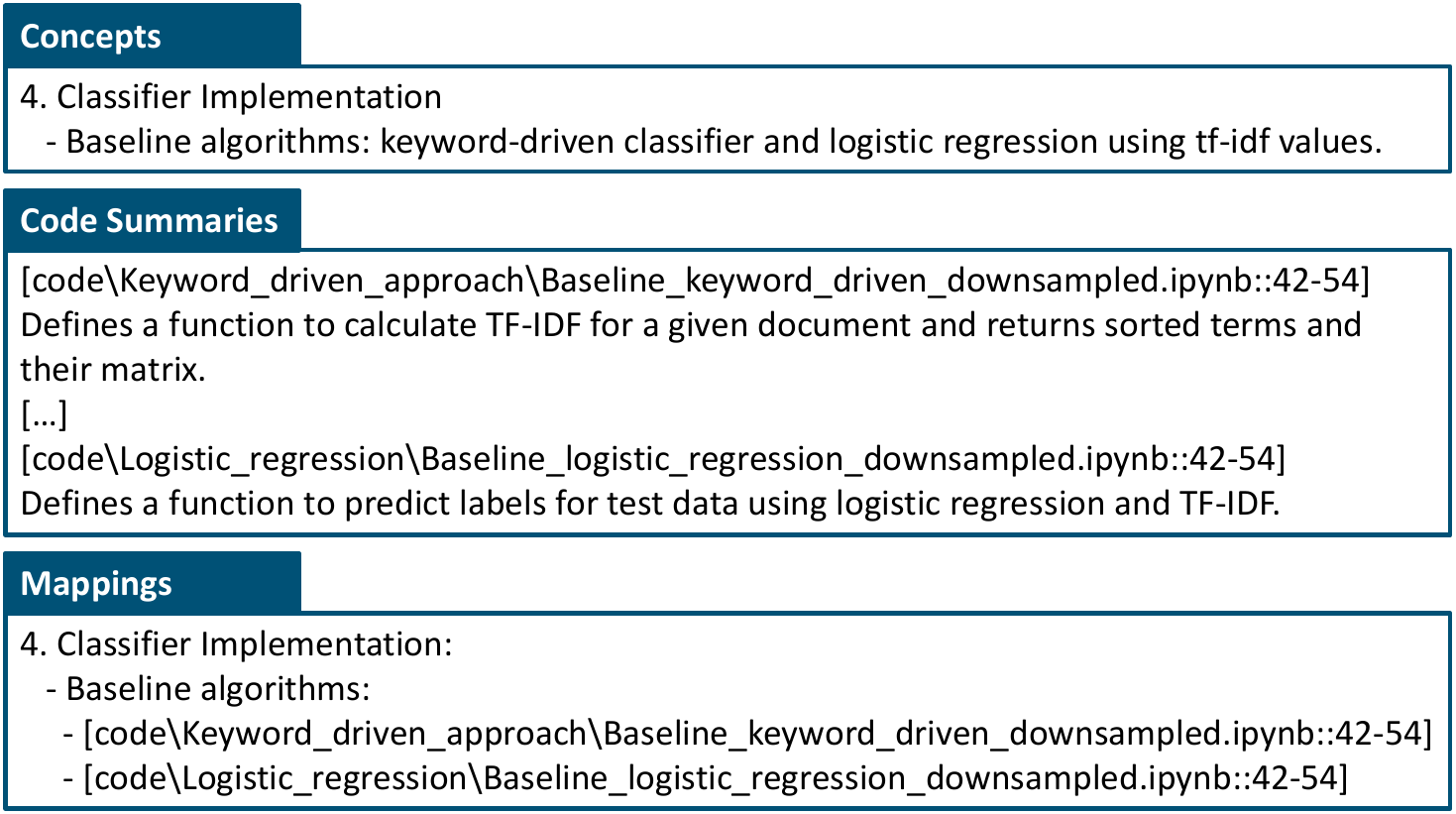}
    \caption{Excerpt from one of the experiments for one particular 3-tuple of concept, code summary, and mapping.}
    \label{fig:placeholder}
\end{figure}

\section{Preliminary Evaluation}
We selected five research groups from different scientific disciplines that actively operate at least one larger research software project. From each, we selected five recent publications, totaling 25 test runs of our approach. In each iteration, we provide the respective paper as a PDF and the project's GitHub link to the tool. We collect the artifacts from each run (code summaries, concepts, and mappings). To assess the quality of the results, we contacted the paper's authors and plan to supply each with the artifacts for their paper to review them with respect to the following key observation criteria:

\textbf{Does code summarization accurately reflect the code's intentions?} (per code block). Possible challenges: too coarse-grained descriptions; misinterpretations of code; hallucinations by the LLM.

\textbf{Did the concept extractor extract all relevant research concepts?} Possible challenges: Missing relevant concepts; irrelevant concepts.

\textbf{How well did the LLM trace the concepts to code?} Possible challenges: Code not related to the concept; concept not implemented; wrong tracing.

Afterwards, we will conduct semi-open interviews with the experts, using open-ended questions, to further assess the quality of the results. We also discuss their ideas for how the tool could assist them in their research.

\subsection{Preliminary Experiences and Discussion}
We conducted initial experiments using familiar research software, e.g., from our own catalog. These primarily stem from the field of software engineering research. The results showed that the tool can extract the main research concepts from the paper, but still considers parts of the paper as main concepts that do not really contribute to the paper's goals, e.g., some introductory definitions. It can summarize code in natural language, concisely and in accordance with a predefined format. It is also capable of recognizing the scientific construct behind an implementation, even when there are no comments indicating precisely what the code implements. It mapped research concepts to code coherently and identified concepts mentioned in the paper but not implemented in code, as well as code blocks that seem essential to the research context but are not mentioned in the paper.

\subsection{Limitations}
The current version of the tool has certain limitations, which we want to run through briefly.
Currently, the tool uses NLRs rather than the code itself for mapping, which might leave the LLM with less information during the mapping process. 

Code that is not encapsulated in a structure (e.g., a method) is grouped into coherent code blocks by the LLM. Thus, the expressiveness of this grouping heavily depends on the LLM's capabilities. 

During code summarization, large code bases are split into chunks that each fit the LLM's context window. This might make the results less accurate in cases where code from another chunk is used,e.g., a method call. In these cases, the LLM has to "guess" what the missing code does.

\section{Conclusion and Future Work}
In this paper, we propose a novel approach to making research software more understandable and transparent to others via an LLM-based tool. It is far from finished, and there are plenty of possibilities to proceed. Beyond the interview study, we aim to incorporate the following aspects:
\begin{compactitem}
    \item A quantitative study in the form of an extensive survey on the quality of the tool's results among a large base of researchers. We aim for 200--300 participants.
    \item A quantitative study over a large number of projects on general data of the tool, e.g., average number of extracted concepts, number of trace links found, size of NL code summaries, etc.
    \item Include a RAG system to get more accurate results.
    \item Extend the tool further for different use cases beyond the scope discussed in this paper.
\end{compactitem}

\bibliography{acmart}
\end{document}